\documentclass[prb,preprint]{revtex4-1} 

\usepackage{amsmath}  
\usepackage{amsfonts} 
\usepackage{graphicx} 

\begin{document}

\title{``Everyone is new to this'': Student reflections on different aspects of online learning}


\author{Danny Doucette}
\email{he/him/his, danny.doucette@pitt.edu}
\author{Sonja Cwik}
\author{Chandralekha Singh}
\affiliation{Department of Physics and Astronomy, University of Pittsburgh, Pittsburgh, PA, 15260, USA\\ 
}

\date{\today}

\begin{abstract}

In 2020, many instructors and students at colleges and universities were thrust into an unprecedented situation as a result of the Covid-19 pandemic disruptions. Even though they typically engage in in-person teaching and learning in brick and mortar classrooms, remote instruction was the only possibility.  Many instructors at our institution who had to switch from in-person to remote instruction without any notice earlier in the year worked extremely hard to design and teach online courses to support their students  during the second half of 2020. Since different instructors chose different pedagogical approaches for remote instruction, students taking multiple remote classes simultaneously experienced a variety of instructional strategies. We present an analysis of students' perceptions of remote learning in their lecture-based, active learning, and lab physics classes at a large research university in the USA, focusing on positive and negative aspects including collaboration, communication, and assessment. Student reflections emphasized the importance of grade incentives for out-of-class and in-class work; frequent, low-stakes assessments; community-building activities; and opportunities to study with peers. Reflecting on the challenges and successes of different types of remote instructional approaches from students' perspective could provide useful insight to guide the design of future online courses as well as some aspects of in-person courses. 
\end{abstract}

\maketitle

\section{Introduction}

In 2020, colleges and universities around the world pivoted to remote instruction as a way to provide educational continuity for learners who are able to attend remote classes during the Covid-19 pandemic disruption. Early reports focused on the impact on teaching and learning that arose from the transition to emergency remote instruction in the northern hemisphere spring of 2020.~\cite{Dew,Vignal,Gavrin,Carleschi,Klein} The switch to remote instruction in the spring of 2020 was hasty, giving instructors no time to develop or cultivate online resources for students, design effective learning experiences, or address the specific opportunities and constraints associated with online learning.

However, by the start of the fall semester, many instructors made heroic efforts to acquaint themselves with the challenges of supporting student learning online, the capabilities of modern technology in delivering remote instruction, and resources and curricular materials available to support instruction.~\cite{Bodegom,Rosen,Campari,DeVore,OBrien,33505} Given the current, ongoing, need for remote instruction because of the pandemic, and anticipating that some colleges and universities may increase the fraction of courses they offer online in coming years, it may be productive to reflect on what worked, and what didn't, from students' perspective on online classes in the fall of 2020. Unlike in the spring of 2020, instructors in fall 2020 invested time and effort designing their classes to be offered online, employing a wide variety of instructional, collaborative, and assessment strategies. We seek here to focus on students' perspectives on remote instruction in a variety of physics courses, as this provides a complement to other studies that focused on student outcomes and instructor perspectives.

At our institution, a large research university in the USA, fall 2020 instruction was fully remote at the start and end of the semester, with a period during the middle when instructors were able to conduct their classes in a hybrid format if they wished. Few instructors employed hybrid instruction during the weeks when it was possible and, for those who did, only a very small number of students attended in-person. In this paper, we analyze survey responses from 1145 students in physics courses at all levels (part of a pre/post survey our department administers for departmental assessment) and interviews with 37 students in physics courses in fall 2020 in order to understand students’ perceptions of the opportunities and limitations of online college physics instruction at our institution. The interviews also helped us understand online science instruction, in general, at our institution. The surveys were given to students in 12 physics courses with 10 different instructors, with an overall response rate of 80\% and response rates from individual classes ranging from 63\% to 100\%. Specifics of the surveyed population are noted in Table~\ref{DemTable}. We expect that our results should generalize to other large research universities in the USA and to other colleges and universities with large-enrollment physics classes, and that many specific topics will resonate with physics instructors in other post-secondary STEM contexts.

\begin{table}[h]
    \centering
    \begin{tabular}{rcc}
       & Surveys & Interviews \\
     \hline
        Introductory Physics & 1109 & 33\\
        Beyond First Year Classes & 18 & 0\\
        Graduate Classes & 18 & 4\\
    \end{tabular}
    \caption{Demographics of students in this study.}
    \label{DemTable}
\end{table}

The surveys included Lickert-scale items that queried students' opinions on whether they were able to have meaningful interactions with peers and instructors, how they felt remote learning compared with in-person learning, and their grade expectations. These items turned out not to be particularly informative since we did not have responses from before the pandemic for comparison. The surveys also included free response questions that asked students about positive and negative aspects of online learning, as well as about what could be done to improve remote instruction. The rich responses that students made to the free response prompts were coded using an emergent concept coding scheme,~\cite{Saldana} and interpreted primarily by considering the frequency of different codes.

Hour-long semi-structured interviews~\cite{Otero} were conducted by all three authors, focusing on students from introductory lecture and lab classes, as well as graduate students. The interviews were transcribed and coded for emergent topics separately from the surveys. Finally, we combined our codes from the surveys and interviews, identifying and reporting only those topics that emerged from both sides of this study. Most topics were similar between different types of classes (e.g., introductory, beyond first year, and graduate), with some exceptions, noted below. The five emergent topics are presented and elaborated upon in the next section: lecture-based classes, active learning classes, lab classes, community and collaboration, and assessment.

As we collected and analyzed these data, we recognized the ethics of teaching and reflecting on student learning during the pandemic. Many of our students have been severely adversely affected by the pandemic in many ways. In fact, the pandemic has brought out the inequities in higher education clearly. Many students have had health issues or have lost loved ones, and others have faced challenging circumstances or paused their studies. We seek to be conscientious about the needs of students who are not represented in our data due to hardships caused by the pandemic, about the additional serious challenges facing students who continued their studies, and about inequities related to who was able to continue their studies. We focus on students' perceptions of online learning rather than potentially biased measures of instructional effectiveness. Our survey data were collected as part of a routine departmental assessment that we conduct every semester. The interviews were conducted with students who were paid for their time. Both survey and interviews were part of IRB-approved studies to improve instruction. The data collection and analysis were originally conducted to provide guidance to instructors in our department on how to tune their online physics courses for the Spring 2021 semester, and the following analysis was undertaken only after we finished providing that guidance.
 
\section{Lecture-Based Classes}
 
In interviews and written survey responses, students expressed both satisfaction and frustration with how the lecture portion of their science courses had been converted into an online format. Among the positives, they described practical advantages to attending lectures online. They could see the slides or whiteboard better and hear the instructor more clearly. A few students noted that instructors made special efforts to clarify some issues with supplementary materials. Studying from the comfort of home was a benefit, as well. Quite a few students described being better able to maintain a restful sleep schedule because they didn’t need to wake up early to commute to class. They felt more physically comfortable connecting to class from the furniture in their homes, and less anxious because they weren’t (hyper)visible~\cite{Reddy} to their classmates. They also reported being more likely to attend class, and more likely to be on time when they did.

\begin{figure}[h]
\includegraphics[width=\columnwidth]{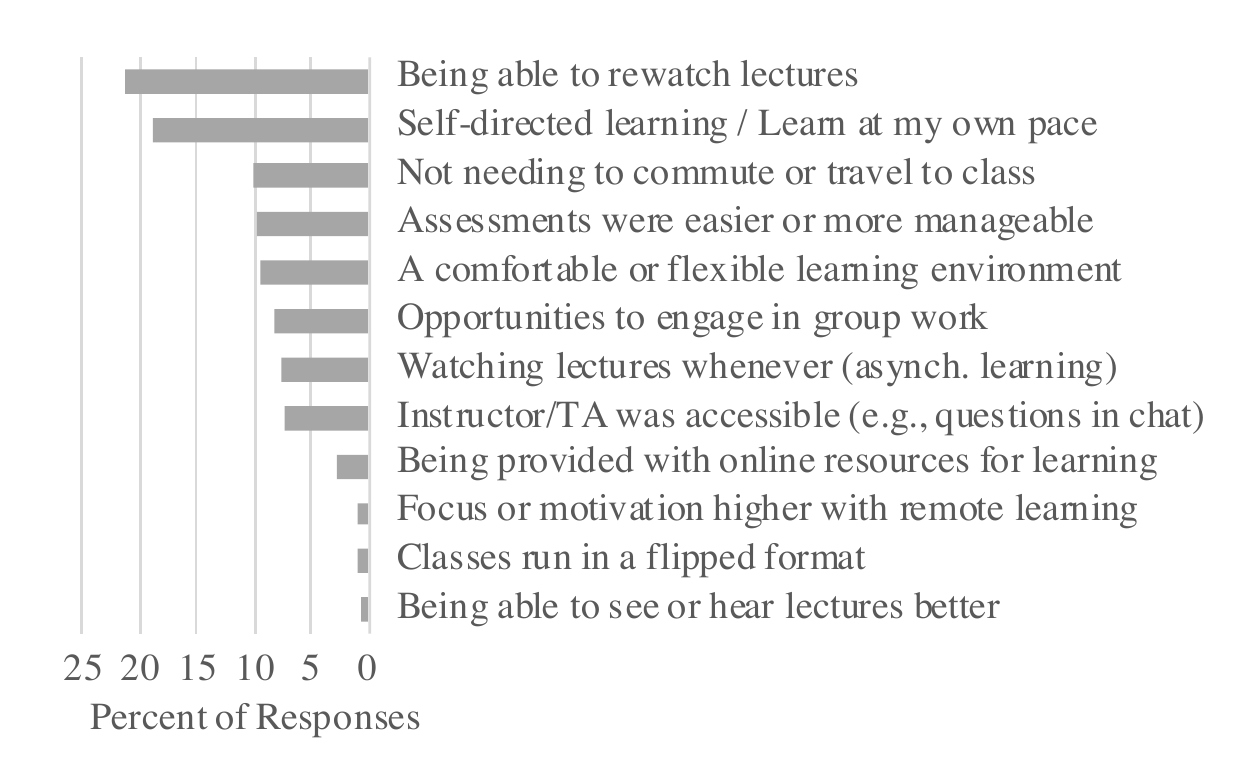}
\caption{Categorized student responses to the prompt, ``Reflecting on the transition to remote instruction, what was the \textit{most} positive outcome?'' Numbers indicate the percentage of students who responded. Some of these categories are related, but we decided to keep them separate since they convey slightly different ideas.\label{Positives}}
\end{figure}

The biggest advantage to live online lectures, according to the students who responded to our survey, was that they could re-watch the lectures (and rewind some parts several times to understand concepts) in order to catch up on the concepts they missed, refresh on ideas later in the semester, or fill in gaps in their notes. In response to a survey question that asked students to identify the single \textit{most} positive outcome of remote instruction (see Fig.~\ref{Positives}), 21\% noted the ability to re-watch live lectures. For example, one student appreciated ``the ability to [go] back and look through the recorded lectures when I'm having difficulty understanding the concepts or need future clarification.''

\begin{figure}[h]
\includegraphics[width=\columnwidth]{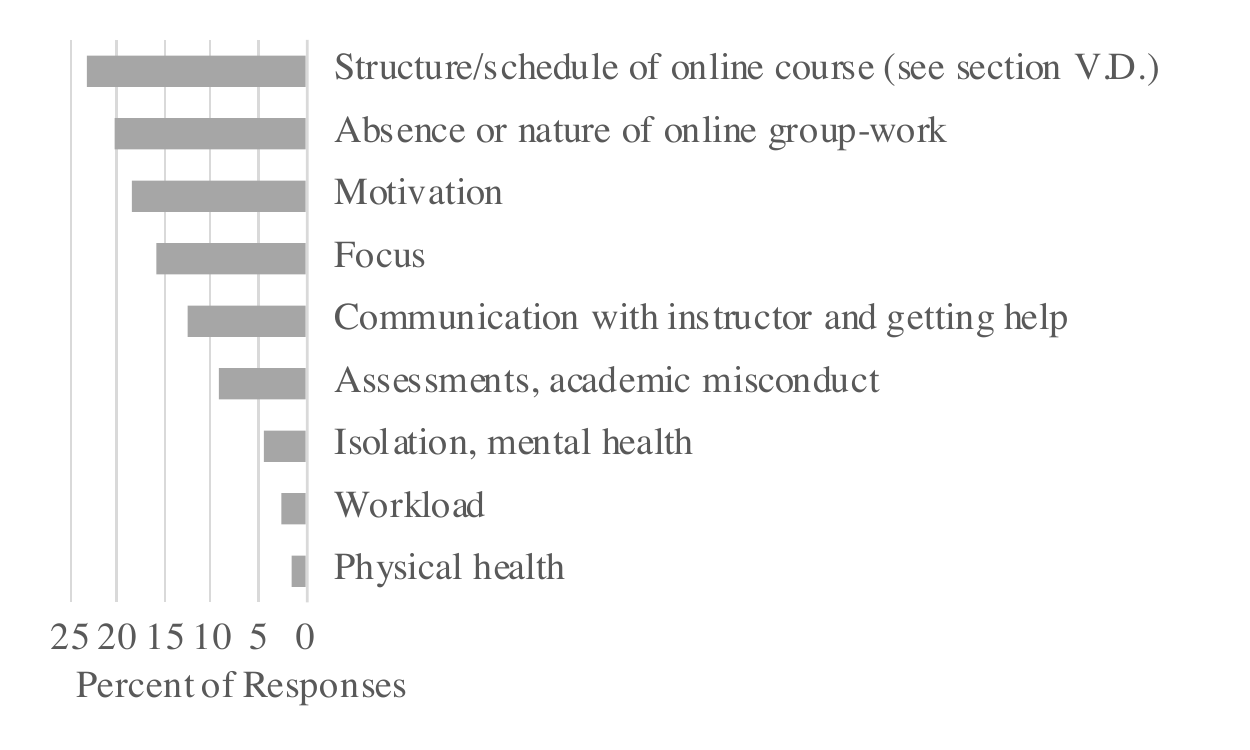}
\caption{Categorized student responses to the prompt, ``Reflecting on the transition to remote instruction, what was the \textit{most} negative outcome?'' Numbers indicate the percentage of students who responded.\label{Negatives}}
\end{figure}

However, the practical benefits of online lectures were balanced with substantial drawbacks. The most significant effect described by students in our survey was a decrease in their focus and motivation because their class was online. While home is comfortable, it is also full of distractions. One student noted, ``I definitely pay attention less than when I was in a normal in-person class, so I get less out of each lesson.'' Another student pointed to ``the issue of staying on task. It's difficult being in your room and staying focused on lectures or homework. It's very easy to get sidetracked on your cell phone or by cleaning your room.'' Without social cues to pay attention in class, it was easy for students to lose focus. A third student noted, ``I could never focus on anything I was doing. Being stuck inside and unable to see how my peers worked made me less motivated.'' The feelings of loss of focus and demotivation went hand-in-hand with poor mental health, with students specifically describing feelings of isolation, loneliness, burn-out, and `Zoom fatigue'.~\cite{Bailenson} In response to a question that asked students to identify the \textit{most} negative outcome of remote instruction (see Fig.~\ref{Negatives}), three of the seven most common topics were student concerns about motivation, focus, and their mental health.

While many students articulated both positive and negative aspects of online courses, first-year students in particular seemed negatively impacted by online learning. Since they were new to the university, they found the online environment made it much harder to adapt to college science learning.  Reflecting on a first semester of college spent taking online courses, one student described the experience as feeling ``fake.'' The transition from high school to college is not always easy. In-person instruction may help some students through the transition by providing opportunities for students to collaborate with one another. Thus, with classes moved online, students may have had fewer opportunities to benefit from the informal norm-setting and expectation balancing that happened previously, when students talked with each other before, during, and after class. Looking forward, it might be productive for colleges and universities to think about ways they could provide a more structured transition from high school to college style learning.

While many of the first-year students we interviewed found online lecture courses to be ``not real", ineffective and disappointing, students in their second year and beyond described a balance of advantages and disadvantages to studying their lecture-based science courses online. The technical benefits of online lectures such as being able to hear and see more clearly, not needing to commute, and being able to re-watch lectures were balanced by concerns about motivation, focus, and mental health.

\section{Active Learning}

The most common alternatives to traditional (but online) lecture-based courses during the fall 2020 semester at our institution were flipped classes,~\cite{Karim} which students encountered at introductory, beyond first year, and graduate levels. In flipped classes, students were expected to watch lecture videos before attending Zoom classes. In class, they engaged in collaborative skills practice activities with coaching from the instructor. In some classes, the instructor used the Zoom polling feature or Top Hat~\cite{TopHat} to replace clicker questions, with grade incentives provided for answering questions. In some cases, students were also graded for the correctness of their polling answers. Breakout rooms were used for small group discussions.

However, in many classes, students were given no grade incentives for completing the in-class collaborative activities or out-of-class work such as watching the videos or answering homework questions. In some classes, the solutions to these activities were posted online. In such classes, students reported that the lack of grade incentives meant that they simply stopped doing these activities on time and prioritized other classes. Moreover, a negative feedback loop occurred such that some students who attended class were unprepared for collaborative work, deflating the quality of group skills practice activities, which students quickly began to consider ``useless.'' One student described, ``Working on a worksheet then brings back the problem ... where people don’t want to participate as much. So then, at the end of the day, you’re not really getting group activity help.'' According to many of the students we interviewed, in some classes where weekly in-class and out-of-class assignments did not have an associated grade incentive they attempted to `cram' before mid-term and final exams by watching videos, rushing through practice problems, and browsing over their posted solutions. Cognitive science tells us that this strategy is largely ineffective.~\cite{Mestre} However, in classes where weekly assignments and collaborative work were graded, students continued to participate throughout the semester and felt that they had learned well and were better-prepared for their exams.

In some introductory classes, the instructor sought to provide completely asynchronous instruction for students, re-purposing scheduled classes into drop-in office hour sessions. The students we interviewed found it very difficult to learn in these classes, as the lack of synchronous interaction decreased their capacity to learn and participate in the class. One student commented, ``it was harder to make sure that you were on pace.'' Without incentive to keep up with lecture videos and homework, most students in asynchronous classes skipped office hours and quickly fell behind. A useful comparison can be made to asynchronous Massive Open Online Courses (MOOCs), where typically only a few highly driven students complete the course since there is little motivation from grades or interactions with the instructor or peers to watch the lectures or complete assignments or assessments regularly.~\cite{SinghIHE}

Thus, when weekly activities were not graded, many students in flipped and asynchronous online classes struggled to keep up. They struggled to self-regulate~\cite{Zimmerman} when they did not encounter collaborative work or an explicit grade incentive. It may be wise for instructors to ensure that their online classes include a synchronous component, and that students are incentivized to complete homework, including watching of videos, collaborative work, and activities during class, such as by assigning a small grade incentive to this work. Of course, instructors should be flexible with students who are unable to participate in these activities each week due to challenging situations and grade them using other means.

\section{Lab Courses}

We asked the students in our interviews to describe both their (introductory) physics labs as well as other (introductory and beyond first year) labs they were taking, including those in other science disciplines. One common approach~\cite{Fox} was for instructors to video-record experiments and have students conduct analysis from the data collected in the video-recording. While they appreciated the effort that goes into making such videos and the paucity of alternatives, the students we interviewed were skeptical of this approach, which they felt served to simplify the labs too much and took away the opportunity to participate in hands-on science. One student commented, in reference to this approach to doing labwork, that ``you don’t really do much work and you don’t do much thinking.''

Another popular approach to doing labs online is to use simulations.~\cite{Fox} Although simulations also simplify lab-work, reducing instruments to cartoons, they still provide students opportunities to make decisions and collect their own data. One student commented, ``I really enjoy ... doing all these online simulations. Everything has been so helpful in truly, truly understanding the material to the best of our ability.''

A smaller number of labs found ways for students to collect their own data, either by having students pick up apparatus, rotate through labs one at a time when it was possible to do so, or use household goods as apparatus. While the students we interviewed appreciated the opportunity to do their own experimentation, none of these approaches worked for everyone. Some had difficulty tracking down particular supplies. Some students had difficulty setting up and operating apparatus without in-person support from peers or the instructor. Others found that without that support, the amount of time they spent troubleshooting their apparatus ballooned.

Students in our introductory physics labs used a combination of simulations and hands-on experiments at home using the IOLab system.~\cite{Iolab} However, regardless of the type of lab-work that students conducted, they felt that collaborative work was an essential part of the lab experience. Students commented that labs that did not include collaborative work felt incomplete. Reflecting on labs that did include collaboration, one student commented, ``I think collaborative activities with your lab group is very important. I think it’s very important that we still do them.'' Many students also missed the in-person aspect of the labs in which they shared the experimentation with other students in a shared space.

\section{Collaboration and Communication}

\subsection{Community and Studying Outside of Class}

Most of the students we heard from via surveys and interviews expressed a desire to collaborate with one another, both in the classroom and out of the classroom in study groups, but faced difficulties doing so. This was common among introductory, beyond first year, and graduate students. When asked about the most negative outcome of online instruction, the most common response described a lack of community engagement and group work. This inhibited students from learning from one another, with one student stating, ``I found it difficult to find help because I do not know how to interact with people I haven't met or know what they look like.'' The lack of collaboration impacted their learning of the material as well as their motivation. It was harder for students to form natural connections with their classmates. Another student explained, ``I miss being able to discuss physics problems and exams with classmates while walking back from class.'' 

Although many students found it harder to work together during the semester, they found benefits from collaborating with their peers when they were able. For example, when asked about the most positive part of online classes, one student mentioned that, ``working on physics problems in a group made me enjoy physics more and really helped me learn.'' Since most of the students we heard from weren’t able to work together in-person, they found new ways to help each other and provide some of the peer support they missed because of online instruction. Another student stated that one benefit of remote instruction was ``having more online groups for each of my classes where students support and motivate each other.'' It could be valuable in future in-person courses for instructors to provide an online environment in which students are given tools and formats to support one another. 

\subsection{Breakout Rooms}

Some professors at our institution encouraged collaboration in their classes by having students work together in video-conference breakout rooms. The students we interviewed described how the structure and group composition of breakout rooms played an important role in determining their success, with several students describing how other members of their group did not participate in breakout room discussions. One student noted that, ``I think over Zoom it did make it a little difficult because nobody really wants to turn their camera on and talk in general.'' Another student talked about the difficulties of working together over Zoom, stating, ``even in recitation, it was difficult to work on assignments together. We all pretty much did our own work and barely talked problems out.'' Therefore, students ended up working alone on assignments that were supposed to involve collaboration.

In our interviews, students also talked about situations in breakout rooms where group work was successful. Several mentioned that a grade incentive for group work was an important factor. For one student in the physics lab, the activities were successful ``because the collaborative activities were mandatory to do and they were graded. So people came to them and worked on them.'' Other students described being able to have successful conversations in breakout rooms if they needed to share an answer with the class when they were done. Structure and incentivized preparation for breakout sessions seem to have been essential for science courses, but less important for non-science courses in which students felt they could participate in conversations even if they hadn't completed the out of class preparatory work. One student described breakout rooms in a world religions class as more talkative and easy to participate in without having done the class preparation. In that world religions class, the breakout room had productive discussions, so the instructor required students to share with the class afterward. The student explained, ``we actually have to come up with an answer and then someone has to share it. I feel like the fear of getting called on and not having an answer prepared makes people talk a little bit more.''

Another advantage of breakout rooms is that there was no noise from other groups talking at the same time. Another disadvantage is that it is difficult for the instructor to know which group needs ``nudging" or help, unlike an in-person class where it is easy for instructors to notice the groups that are productively engaged in discussion and the ones that are not.

Along with a grade incentive, teaching assistants (TAs), undergraduate teaching assistants (UTAs), or undergraduate Learning Assistants (LAs) could play a valuable role in stimulating small group discussions. One student explained, ``I think a big motivator is when the TA does come around to the breakout rooms for people to speak up.'' TAs might be trained and provided with a `script' of example questions they could ask to help steer the discussion in a productive way. When they were structured, incentivized, and supported effectively, students typically reported that breakout rooms were a valuable and productive aspect of their online learning.

\subsection{Back Channel Chat}

A back channel is a way in which students can ask questions or seek clarification from the instructor during class, without interrupting the main flow of the lesson by, e.g., raising a hand and posing a question. It can be challenging for students to raise their hands and ask questions in class, knowing that by speaking up they are subjecting themselves to the judgment of their peers. The back channel reduces this fear. One student noted, ``I don't feel shy to ask questions because I'm not in-person.'' According to the students we interviewed, some instructors were able to use the chat feature in Zoom to provide students with a back channel to ask questions during class. In some large classes with an assigned TA, the instructor would task the TA with monitoring the chat, responding to questions when possible, and flagging some questions for the instructor to answer. Sometimes, however, the back channel became dominated by a small number of personalities, and some students reported being so engrossed in the discussion on the chat that they were no longer paying attention to the lecture. Productive uses of the chat as a back channel typically required the establishment of norms for the use of the chat. When in-person classes resume, it may be worthwhile exploring how in-person lectures could incorporate recording and back channels to support student learning. It may be especially valuable to have a TA monitor the chat, if possible.

\subsection{Office Hours and Communication}

Most of the students we heard from found that online office hours were easier to attend than in-person meetings. In addition to not needing to walk over to the physics department, several students said that they appreciated being able to log in to office hours with their camera off and listen to other students' questions if they weren't yet ready to ask their own. However, several other students felt that online office hours were more difficult to attend. For example, one student in a large-enrollment gateway science class noted that it was ``harder to ‘pop by’ the office for a quick question.'' In a few cases, instructors of beyond first year and graduate courses asked students to contact them to schedule an office hour visit. This approach seemed to deter students from seeking extra help, as it made the office hour feel more formal and removed the possibility for students to listen in with their cameras off until they felt comfortable participating actively.

Several students found it more difficult to ask questions or get feedback from their instructors; difficulty in communicating with their instructors was a common theme in the survey responses. A few students commented that their instructors didn't always reply to their emails. According to one student, ``I completely rely on my professor's validation and comments to know how well I'm doing in this class. I couldn't get any of that because I couldn't interact with my professors.'' Virtual drop-in tutoring was perceived to be less effective than in-person drop-in tutoring. 

In survey responses, many students described struggling to understand the structure or requirements of their courses, especially when multiple online tools were used. One student disliked ``checking multiple platforms for assignments and potentially missing assignments because of it.'' According to the students we heard from, some instructors also struggled with technology. While many instructors were applauded for introducing new digital tools or using standard online tools effectively, in other cases students felt that considerable instructional time was lost by instructors who struggled with the technology. Considering the short time instructors had to learn to use these tools, some difficulty with technology usage was inevitable.

The question of whether students should have their cameras on or off came up frequently in our interviews and survey responses. While most students understood that instructors preferred to lecture to classes that had their cameras on, they also expressed a variety of reasons why they sometimes preferred to keep their cameras off. These included concerns about internet bandwidth, appearing on-camera in their pajamas, and other issues related to attending class from home. However, the majority of the students we interviewed preferred that their peers turn their cameras on for breakout room discussions, and expressed frustration when their classmates did not do this. It might be productive for instructors to set clear expectations for camera usage in their classes, such as strongly recommending but not requiring their use in breakout rooms.

In all types of classes, the students we interviewed appreciated that some of their instructors started their classes with a check-in to see how everyone was doing. They also shared their own adversities while living through the pandemic as well as things that brought them joy (such as their pets). In written survey responses, students wrote that they ``appreciated [professor's] positivity and desire to still help us learn'', that they ``feel as though I got to work with a professor who genuinely cares about her students'', and that ``many of my professors were accommodating to the circumstances, which made things less stressful.''

\section{Assessment}

The students we surveyed and interviewed expressed a wide variety of opinions about assessment in online courses. Many found online quizzes to be easier, online tests written at home to be less stressful, and open-note exams to provide a better opportunity to demonstrate their understanding of physics concepts. Others disliked short, timed quizzes; struggled with scanning and uploading work; and felt that the assessments were more challenging than they had been in-person. A small number of students expressed enthusiasm for video-based assessments that either required them to record short videos in which they solved problems or that involved a short, low-stakes oral examination via Zoom.

Overall, the majority students in introductory, beyond first year, and graduate classes preferred a strategy of frequent, low-stakes assessments. Frequent assessments provide plenty of feedback, low stakes keep anxiety low, and the flexibility inherent in frequent low-stakes assessments can help students. By decreasing anxiety, frequent low-stakes assessments may make introductory STEM courses more equitable.~\cite{Ballen,Cotner} A few students expressed anxiety related to feeling uncertain about shifting grading policies, but they were generally positive once they understood how they would be assessed.

One physics instructor implemented a group portion to the exams in their introductory class. In this two-stage group exam, students worked in groups of four on typical physics problems. Then, the next class, the students individually wrote responses to questions that asked about the strategies and physics concepts they used to solve the problems. The instructor's goal was to provide students with an opportunity to collaborate with peers during the assessment, while still providing a measure of individual accountability, as a way to decrease the pressure that students might be feeling to engage in academic misconduct during assessments. The order of this group exam is opposite that which is often described in the literature.~\cite{Wieman,Jang} One student shared their experience with the two-stage group exams, stating, ``I felt that working with my group was rather enjoyable and it helped ease some stress I had had about physics in the past.'' Other instructors divided the total points across the semester so that even though the final week's assessment was cumulative, it was not worth as many points as it typically is. Students generally appreciated this approach, in which one exam did not count for too much of their grade. Instructors should consider some of these approaches in their in-person classes as well.

\section{Discussion and Conclusion}
 

While the findings presented in this paper may be informative for instructors of online classes in the future, we believe that some of the lessons from this transition to remote instruction are also relevant to in-person instruction. In flipped classes, it may be important to provide low-stakes grade incentives to make sure students are engaging with all aspects of the work, inside and outside of the classroom. More frequent, low-stakes (formative) assessment, and less weight on a final exam, can support spaced practice and reduce student anxiety. Even if a back channel is not possible in an in-person class, it may be possible to provide students with several opportunities during each lecture to talk with their peers and identify the concepts that are most challenging for them so that instructors can address them during class. Continuing to hold some virtual office hours, leveraging our expertise with technology such as learning management systems, and focusing on clarity in vocal and written communication may also be useful take-away messages from our experiences with remote instruction. Finally, a subtext to much of this work is that instructors and students were more willing to discuss their struggles with teaching and learning with each other during the pandemic. The shared humanity was appreciated by many students during this difficult semester, but as long as college is stressful for students, connecting with students will likely be time and energy well invested. It may be useful for instructors to retain some of that human connectedness and mutuality when we return to in-person instruction.

\begin{acknowledgments}

We wish to recognize the instructors who have gone `above and beyond' to support student learning during these difficult times. We are very grateful to our interview participants and survey respondents for their time, feedback and insight during an extremely challenging semester. We also acknowledge the large number of educators associated with the dB-SERC and AAPT whose insight and ideas, shared via lunch meetings and coffee hours and social media, have informed our own in ways too subtle to be called out specifically. Thank you all.

\end{acknowledgments}

\end{document}